\documentclass[twocolumn]{RAA}            % referee version: for submission

\usepackage{graphicx,times}             %for PS/EPS graphics inclusion, new
\usepackage{natbib}
\usepackage{amssymb,amsmath}
\usepackage{enumerate}
\bibpunct{(}{)}{;}{a}{}{,}
\usepackage{color}
\usepackage[pagebackref=true]{hyperref}

\usepackage{caption}
\usepackage{newtxtext,newtxmath}
\usepackage{subcaption}
\usepackage{ulem}

\newcommand{\Mpc}{\mathrm{Mpc}}
\newcommand{\hMpc}{h^{-1}\mathrm{Mpc}}
\newcommand{\hMass}{h^{-1}\mathrm{M}_\odot}

\begin{document}

  \title{A widely applicable Galaxy Group finder Using Machine Learning
}
%   \subtitle{I. Place Your Subtitle Here}

   \volnopage{Vol.0 (20xx) No.0, 000--000}      %%preserved for Editor. DOn't remove!
   \setcounter{page}{1}          %%starting page, preserved for Editor. DOn't remove!

   \author{Juntao Ma \inst{1,2}
   \and Jie Wang \inst{1, 2}
   \and Tianxiang Mao \inst{1, 2}
   \and Hongxiang Chen \inst{1, 2}
   \and Yuxi Meng \inst{1, 2}
   \and Xiaohu Yang \inst{3, 4}
   \and Qingyang Li \inst{3, 4}
   }

   \institute{National Astronomical Observatories, Chinese Academy of Sciences, Beijing 100101, China; {\it jtma@nao.ac.cn, jie.wang@nao.cas.cn}\\
        \and
             School of Astronomy and Space Science, University of Chinese Academy of Sciences, Beijing 100049, China\\
        \and
             State Key Laboratory of Dark Matter Physics, Tsung-Dao Lee Institute \& School of Physics and Astronomy, Shanghai Jiao Tong University, Shanghai 201210, China\\
        \and 
             Shanghai Key Laboratory for Particle Physics and Cosmology, Shanghai 200240, China\\
\vs\no
   {\small Received 20xx month day; accepted 20xx month day}}

\abstract{ Galaxy groups are essential for studying the distribution of matter on a large scale in redshift surveys and for deciphering the link between galaxy traits and their associated halos. In this work, we propose a widely applicable method for identifying groups through machine learning techniques in real space taking into account the impact of redshift distortion. Our methodology involves two neural networks: one is a classification model for identifying central galaxy groups, and the other is a regression model for predicting the mass of these groups. Both models input observable galaxy traits, allowing future applicability to real survey data. Testing on simulated datasets indicates our method accurately identifies over $92\%$ of groups with $\mathrm{M}_{vir} \geq 10^{11}\hMass$, with $80\%$ achieving a membership completeness of at least $80\%$. The predicted group masses vary by less than 0.3 dex across different mass scales, even in the absence of a priori data. Our network adapts seamlessly to expand to sparse samples with a flux limit of $m_{r} < 14$, to high redshift samples at $z=1.08$, and to galaxy samples from the TNG300 hydrodynamical simulation without further training. Furthermore, the framework can easily adjust to real surveys by training on redshift distorted samples without needing parameter changes. Careful consideration of different observational effects in redshift space makes it promising that this method will be applicable to real galaxy surveys.
\keywords{large-scale structure of Universe -- Galaxy: halo  -- methods: data analysis}
}

   \authorrunning{Juntao M. et al.}            %author_head in even pages
   \titlerunning{A General group finder}  % title_head in odd pages

   \maketitle
%% The author head (on even pages) and the title head (on odd pages) will be
%% automatically extracted from \author{} and \title{}. Whenever the title is too long,
%% you will be asked to supply a shorter one by inserting either \authorrunning{} or
%% \titlerunning{} before \maketitle. Anyway, you can specify your own heads.
%%
%%
%% Note: In the following text body of your manuscript, please note several differences from
%%       other major journals:
%% (1) \subsection{Please Capitalize the First Letter of Each Notional Word in Subsection Title}
%% (2) Please Capitalize the First Letter of Each Notional Word in all tables' captions

%
%________________________________________________ sections below
%
\section{Introduction}

Current structure formation theory suggests that the mass content of the universe is dominated by dark matter, and cosmic structures form hierarchically through gravitational instability \citep{white1978core, davis1985evolution, springel2006largescale}. Galaxies form and evolve within these structures, called dark matter halos. Therefore, understanding the relation between galaxies and their host halos is crucial for understanding the role played by the environment in galaxy formation and evolution, as well as for tracing the underlying density field. Apart from theoretical or simulation-based approaches, galaxy groups provide a direct way to study the galaxy-halo relation, as they consist of various galaxies residing within the same mass dark matter halos.

Due to the visual concentration of these systems, identifying galaxy groups from galaxy surveys begins at the very beginning of cosmology research. \citet{abell1958distribution} identified approximately 2700 clusters from the Palomar Observatory Sky Survey (POSS) using local galaxy surface number densities. Similarly, \citet{zwicky1968catalogue} constructed a catalogue of 9133 clusters in the Northern celestial hemisphere, and \citet{abell1989catalog} identified around 1600 clusters from the UK Schmidt Telescope (UKST) plates. The lack of precise distance estimation in early no-redshift surveys caused these catalogs to suffer significantly from issues like projection effects. With the advent of large redshift surveys since the 1980s, many efforts have been made to identify galaxy groups using different group finder algorithms. For instance, galaxy groups have been identified from the CfA redshift survey \citep[e.g.][]{huchra1982groups}, the Two Degree Field Galaxy Redshift Survey \citep[e.g.][]{eke2004galaxy, yang2005halo, tago2006clusters}, the Two Micron All Sky Redshift Survey \citep[e.g.][]{lavaux20112m++, tully2015galaxy, crook2007groups}, the Sloan Digital Sky Survey \citep[e.g.][]{goto2005velocity, berlind2006percolation, yang2007galaxy, lim17galaxy}, the DESI Legacy Imaging Surveys \citep[e.g.][]{yang_extended_2021}.

Based on the galaxies groups identified from large redshift surveys, we can have a better understanding on how different galaxies form and evolve in different dark matter haloes. \citet{weinmann2006properties} found a strong correlation in the properties of galaxies residing in common dark matter halos, i.e. galactic conformity. \citet{wang2018elucid} found that the apparent dependence of the quenched fraction of galaxies on large-scale environment is largely induced by the dependence of quenching on the host halo mass combined with the biased distribution of dark matter halos in the cosmic density field. The group-galaxy cross-correlation function is measured to evaluate how galaxies are distributed within and beyond their host halos \citep[e.g.][]{yang2005cross, coil2006deep2, knobel2009optical}. Stacking groups with similar masses can help probe the weak signal of Sunyaev-Zel$^,$dovich (SZ) effects \citep[e.g.][]{li2011probing, vikram2017measurement, lim2018gas, lim2020detection} and weak gravitational lensing signals \citep[e.g.][]{mandelbaum2006density, yang2006weak, han2015galaxy, viola2015dark, luo2018galaxy} over a large halo mass range. A similar approach can be used to measure the halo occupation distribution or the conditional luminosity functions of galaxies in halos of different masses \citep[e.g.][]{yang2005halo, yang2008galaxy, yang2009galaxy, rodriguez2015taking, lan2016galaxy}. As biased tracer of dark matter, galaxy groups and their halos can be used to reconstruct cosmic density field \citep{wang2009reconstructing, munoz2011halo} and constrain the initial conditions that produced the observed cosmic web \citep[e.g.][]{wang2016elucid}.

Several group finders have been proposed and applied to redshift surveys. The Friends-of-Friends (FoF) algorithm identifies galaxies as belonging to the same group when their distance is less than a linking length. For example, \citet{huchra1982groups} used a FoF method with two linking lengths, one in the projected direction and the other in the redshift direction, to construct galaxy groups in the CfA survey. \citet{miller2005c4} used the C4 algorithm, which places galaxies in a six-dimensional parameter space, to find groups in the SDSS DR2. \citet{yang2005halo} proposed a halo-based group finder that takes advantage of known halo models (e.g., NFW profile) and iteration.  \citet{wang2020identifying} combined FoF and machine learning methods to identify groups in incomplete samples at high redshift.

Traditional group finders rely on specific physical models that statistically describe the structures of galaxies and halos. These models may lose higher-order information present in the cosmic density field. In this paper, we propose a machine-based group finder algorithm, specifically artificial neural networks (ANNs). Machine learning models learn directly from obervational data, allowing us to extract more nuanced information hidden in the data provided to the networks. Recent research has shown that machine learning models outperform traditional methods in various tasks, such as feature extraction and classification. Our recent research has confirmed the effectiveness of Artificial Neural Networks (ANNs) in solving a variety of astrophysical problems.As an example, \citet{mao_baryon_2021} introduced an innovative convolutional neural network framework for reconstructing baryon acoustic oscillation (BAO) signals, 
significantly enhancing the BAO signal-to-noise ratio to around 
$k \simeq 0.4 h \mathrm{Mpc}^{-1}$.
Similarly, \citet{chen_estimation_2024} utilized ANNs to assess environmental attributes of galaxies,achieving accurate line-of-sight velocity estimations and enabling the recovery of the real-space power spectrum with less than a 5\% margin of error.

 For our group finder, we aim to develop a machine learning model that accurately identifies member galaxies and estimates halo mass. Moreover, with appropriate preprocessing and network architecture, machine learning models can demonstrate good generalisability, meaning they can be applied to different galaxy catalogues without the need for retraining or hyperparameter adjustment. By leveraging ANNs, our group finder extracts information about galaxies and their host halos from a high-resolution N-body simulation. The nonlinear nature of the network enables it to uncover more intricate relations than those described by current galaxy-halo models. Our group finder demonstrates remarkable accuracy across various test datasets, including those at different cosmic epochs, and galaxy samples with different flux limits. 

This paper is organised as follows. In Section 2 we provide a description of the simulation data used for training and testing our group finder. In Section 3 we describe in detail our group finding method, which includes two machine learning models. In Section 4 we test the performance of our group finder, including completeness and purity test, halo mass assignment test, on each of the test datasets. Finally, we conclude our main results in Section 5.

\section{Data}
This section outlines the data sets utilised in our study. The Millennium Simulation is partitioned into a subbox with a side length of 300 Mpc/h for training purposes, as well as several smaller boxes for testing to maintain data integrity. In addition, to assess the robustness and flexibility of our model, we developed three additional test datasets.

\begin{enumerate}
    \item Magnitude Sample: Similar to the basic test datasets, but with a different apparent magnitude limit applied.

    \item High-z Sample: Consisting of galaxies from higher redshift snapshots of the Millennium Simulation.

    \item TNG Sample: Generated from the TNG300 simulation, providing a distinct environment for model evaluation.
\end{enumerate}
These samples provide a diverse range of data for comprehensive testing. Additionally, we evaluated our model on a redshift-space dataset to demonstrate its applicability to real redshift surveys. We detail these data set as follows.

\subsection{Traning Data} \label{mock}
To generate suitable training datasets for our machine learning models and evaluate the performance of our group finder, we utilized cosmological simulation, Millennium simulation, and its galaxy catalogue. 

The Millennium Simulation(MS; \citet{springel2005simulation}), a large-scale simulation of cosmic structure formation based on the $\Lambda$CDM cosmology. It simulates $N = 2160^3$ dark matter particles across a redshift range from $z=127$ to $z=0$, within a co-moving volume of $(500\hMpc)^3$. Each dark matter particle has a mass of $8.6\times 10^8 M_{\odot}$. The cosmological parameters of the simulation are $\Omega_m = 0.25$, $\Omega_b = 0.045$, $h=0.73$, $\Omega_\Lambda = 0.75$, $n=1$, and $\sigma_8=0.9$, with the Hubble constant defined as $H_0=100h\ \mathrm{Km \cdot s^{-1} \cdot \Mpc^{-1}}$.

We use the semi-analytic galaxy catalogue of MS developed by \citet{guo11from}, which implement the galaxy formation model L-Galaxies \citep{henriques2015galaxy}  onto merger trees extracted from the Millennium Simulation.

A sub-box with dimensions of $300 \times 300 \times 500(\hMpc)^3$ from the MS was selected as the training set. Edge effects of the box may lead to some groups being incomplete in terms of member galaxies, so these incomplete groups were removed in data pre-processing. Furthermore, we selected only groups with host halos containing more than 100 dark matter particles. The observer is at edge plane of the simulation box, with the perpendicular axis to this plane as the line-of-sight direction. The apparent magnitudes of galaxies were computed based on their absolute magnitudes and their line-of-sight distances, adhering to an r-band magnitude limit of $mag_r < 17.7$, which aligns with observational sample criteria of the SDSS sample \citep{sdss2009}. The resulting galaxy catalogs formed the foundation for training our machine-learning models. The training dataset includes 1,298,413 galaxies distributed in 945,078 dark matter halos.

\subsection{Test Data} \label{test}

The basic test data sets were generated from six small sub-boxes in the MS, each with a size of $(200\hMpc)^3$. All the selection criteria were same as those used for the training datasets. The resulting six basic test datasets contain a total of 1,650,251 galaxies distributed across 1,189,865 dark matter halos.

We also evaluated our model using three extended datasets. The magnitude limited datasets, derived from the same simulation box as the training and basic test datasets, include different apparent magnitude limits for the r-band: $mag_r < 16$, $mag_r < 15$, and $mag_r < 14$, resulting in galaxy catalogues containing 169,375, 118,931, and 77,950 galaxies, respectively. The high-z datasets comprise galaxies from higher redshift snapshots at $z = 0.32$ (Snapshot 52), $z = 0.62$ (Snapshot 46), and $z = 1.08$ (Snapshot 40) of the Millennium Simulation, using the same limit of apparent magnitude as $mag_r < 17.7$. These high-z datasets contain 593,102, 651,365, and 670,209 galaxies, respectively.

The TNG samples are derived from the IllustrisTNG Project, which builds upon the earlier Illustris simulation. TNG features cosmological magnetohydrodynamical simulations aimed at understanding key mechanisms in galaxy formation and evolution. It includes three main simulation runs with different scales and resolutions: TNG50, TNG100, and TNG300. In our research, we use TNG300, which is the largest simulation, which includes a volume of $(302.6\Mpc)^3$. This simulation begins at redshift z=127 and is based on the Planck 2015 $\Lambda$CDM cosmological parameters ($\Omega_{\Lambda,0}=0.6911$, $\Omega_{m,0}=0.3089$, $\Omega_{b,0}=0.0486$, $\sigma_8=0.8159$, $n_s=0.9667$, $h=0.6774$), with dark matter particles having a resolution of $m_{DM} = 5.9\times 10^7 M_\odot$ and gas cells averaging $m_{gas} = 1.1 \times 10^7 M_\odot$ in mass. Our study selects galaxies within subhalos containing over 20 dark-matter particles and positive stellar mass ($M_s > 0$). We apply apparent magnitude cut-off of $mag_r \leq 17.7$, using z = 0 as the observer's reference frame. The TNG300 galaxy mock catalogue includes 338,161 galaxies and 221,971 groups.

To assess how well the model applies to actual redshift surveys, we evaluated it using a redshift-space mock catalogue. Despite the fact that the machine learning model was initially trained on non-redshift-space data, our group finder does not depend on exact redshift measurements, making it adaptable to redshift space. The redshift distorted samples is sourced from the $z = 0$ snapshot of the MS. As with the training data, groups intersecting the box boundaries and those with host halos having fewer than 100 dark matter particles were omitted. In this dataset, galaxy line-of-sight distances were adjusted factoring in redshift distortion effects, determined by their line-of-sight velocities. An apparent magnitude limit of $mag_r < 17.7$ was maintained. The redshift distorted samples comprises 397,283 galaxies in 293,030 groups.

Same as the training data, the observer of all test simulation boxes will also be placed at edge plane of the simulation box, and its perpendicular axis as the line-of-sight direction.

\section{Method}

Our group finding algorithm is based on artificial neural networks (ANN). ANNs are a rapidly growing area of machine learning, and many network architectures, such as multilayer perceptrons, convolutional neural networks, graph neural networks, and recurrent neural networks, have been developed to solve different types of problems. ANNs use non-linear models to solve complex problems by optimising trainable parameters through gradient descent. This training process involves feeding data to the network to adjust these parameters.
 
Our group identification system utilizes observable attributes of galaxies to detect groups and predict their virial masses. The framework is comprised of two machine learning models: 1) Central Galaxy Identifier, a classification model that identifies the most likely central galaxy from a galaxy's nearby neighbors; 2) Group Mass Estimator, a regression model that calculates a group's mass based on its member galaxies' properties. The initial output from the central galaxy identifier generally results in smaller group segments that align with actual groups, which are regarded as group candidates. These candidates undergo merging according to certain criteria, considering their virial mass projections from the group mass estimator and their spatial arrangement. This merging process continues until the group catalog stabilizes with no further modifications. A thorough description of these models and ensuing steps will be discussed in the next sections.

\subsection{Identification of Central Galaxies}

The catalog of galaxy groups can be described by the center-satellite systems, which highlights the importance of pinpointing the central galaxy for every group. To facilitate this, we have created an artificial neural network (ANN) to recognize central galaxies by analyzing their environmental features. The network uses the following properties as input:

\begin{align*}
&M_{r, 0}: \text{r-band magnitude of target galaxy} \\
&(g-r)_0: \text{color of target galaxy} \\ 
&z_0: \text{redshift of target galaxy} \\
&d_i: \text{projection distance to its neighbors} \\
&M{r, i}: \text{r-band magnitude of neighbor galaxies} \\
&dz_i: \text{redshift distance to its neighbors} \\
&(g-r)_i: \text{color of neighbor galaxies}
\end{align*}

Here, the target galaxy refers to the galaxy for which we want to identify the central galaxy. Central galaxies are selected from the 10 nearest neighboring galaxies of the target galaxy, with $i$ representing the rank of the neighbors, ranging from 1 to 10. When analyzing actual survey data, the redshift distortion will cause uncertainties in the estimation of line-of-sight distances. To account for this, we allow a $\pm 5 \hMpc$ margin when pinpointing nearby galaxies. Galaxies positioned within $\pm 5 \hMpc$ of a target galaxy's line-of-sight are deemed equivalent, and only projection distances are used to identify such neighbors. These neighboring galaxies are ranked according to their projection distance to the target galaxy. In the simulation, we verified that the central galaxies of approximately $94\%$ of the galaxies lie within their 10 nearest neighbors plus themselves.

Target galaxies are categorized into 12 distinct classes based on their identification outcomes. Specifically, the ten closest neighbors are numbered from 1 to 10, ordered by their projected distances ($d_i$) in ascending sequence, and the target galaxy is indexed as 0. If the central galaxy of the target galaxy is included in these 11 galaxies, it is given the label corresponding to the central's own index. If none of these galaxies serve as its central galaxy, it receives a label of 11.

Our neural network consists of four hidden layers, each with a Rectified Linear Unit (ReLU) activation function. The network outputs a 12-element vector representing the probabilities of a galaxy belonging to each of the possible classes. We use cross-entropy as the loss function and train the neural network using the training dataset from the Millennium Simulation (MS) for 500 epochs.

\begin{figure}
    \centering
    \includegraphics[width=\columnwidth]{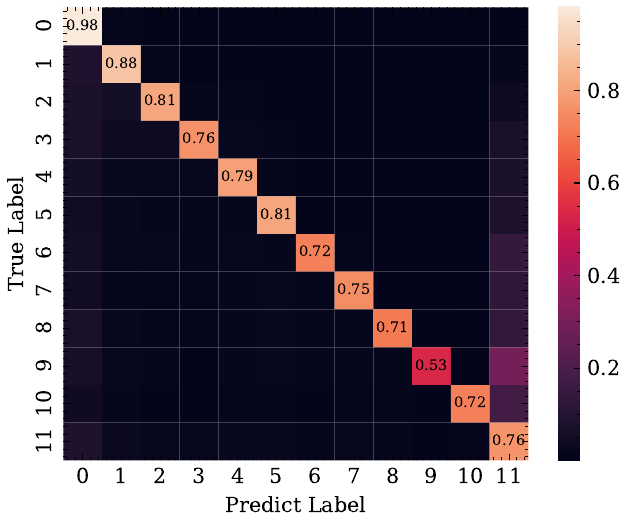}
    \caption{ Confusion matrix for central galaxy identification on the test dataset. The x-axis represents the predicted labels, while the y-axis represents the true value. Each number within a cell represents the fraction of galaxies with a true label of 'y' that were predicted as label 'x.' Consequently, the sum of the numbers in each row equals 1. The color shading within the matrix also reflects this fraction.
    }
    \label{fig:confusion Matrix}
\end{figure}

Figure \ref{fig:confusion Matrix} illustrates the confusion matrix derived from our findings. A confusion matrix (or error matrix) offers a comprehensive overview of accurate and inaccurate classifications. Notably, the class designated as 0, which identifies the galaxy itself as the central galaxy, is the most prevalent and exhibits the greatest accuracy. The other classes maintain acceptable accuracy, although there is a declining trend as the class label numbers increase. Additionally, it is significant to note that a considerable fraction of other classes is classified as class 11.

Although the identifier for central galaxies demonstrates strong accuracy in various galaxy samples, there remain cases where galaxies are inaccurately categorized as central, or the central counterpart is not found (assigned to class 11). Importantly, the core objective of the central galaxy identifier, as an initial component of our group finder, is not to accurately assign a central galaxy to each individual galaxy. Instead, it aims to ensure that the predicted central-satellite pairs are situated within the same dark matter halo, thus making them viable candidates for groups in subsequent merging processes.

Furthermore, with the test data, it is observed that 99.02\% of the center-satellite pairs coexist within the same halo. This suggests that the groups formed by the central galaxy identifier serve as an effective initial step in the identification of galaxy groups.

\subsection{Estimation of group mass}

The host halo mass of galaxies is vital for comprehending the development and dynamics of galaxies. In order to approximate the halo mass of the galaxy group identified in the previous section, we developed a new artificial neural network specifically designed to forecast the halo mass of groups. This is accomplished by using data from the central galaxies and the it's top large satellite galaxies within these groups.

The network takes the following properties of a target group as inputs to predict its halo mass:
\begin{align*}
&M_{*,c}: \text{stellar mass of central galaxy} \\
&M_{*, t}: \text{total stellar mass of all galaxies in the group} \\
&N: \text{total number of galaxies in the group} \\
&M_{r, max}: \text{maximum of r-band magnitude among all galaxies} \\
&M_{*,i}: \text{stellar mass of most massive 5 satellite galaxies} \\
&d_{i}: \text{projection distance of the 5 satellites to group center}
\end{align*}

The variable $i$ represents the index of five most massive satellite galaxies within the group, $i=1$ is the most massive satellite galaxy. In cases where a group has fewer than five members, $M_{*, i}$ and $d_i$ will be filled with zeros.

The network consists of four hidden layers, each using a Rectified Linear Unit (ReLU) activation function. We use Mean Squared Error as the loss function and train the neural network on training dataset from the Millennium Simulation (MS) for 500 epochs.

\subsection{Identification of groups} \label{sec:correction}

Although the central galaxy identifier shows commendable precision with simulated data, deeper scrutiny discloses that it may unintentionally divide larger groups into smaller subsections. This division can adversely affect the halo mass function and hinder the accurate identification of massive halos. To remedy this, we suggest an iterative method to reconcile and unify sections of authentic groups, commencing with those identified by the central galaxy identifier.

The correction methodology is outlined in the steps below:

\begin{enumerate}

    \item\label{gf:step1} Group and Mass Estimation: The groups predicted by the machine learning algorithm are considered as candidates. The Group Mass Estimator is employed to forecast the halo mass, labeled as $M_{200}$, for each group. We then compute $R_{200}$ using the equation ${M_{200}}/{\frac{4}{3}\pi r_{200}^3} = \rho_c$.
    
    \item\label{gf:step2} Group Consolidation: For every candidate group, all galaxies located within its $R_{200}$ radius are examined. As mentioned before, we also set a line-of-sight $\pm 5\hMpc$ tolerance when searching for these near galaxies, which will guarantee the ability of our model to be further used in redshift surveys. If neighboring galaxies belong to another group, the two groups are combined into one. The new group's center is aligned with the more massive of the initial groups, determined by their predicted halo masses, integrating all members from the original groups. 
    
    \item\label{gf:step3} Reiteration: After merging, an updated group catalog is created. These revised groups are used as candidates to predict new halo masses, repeating the second step until the group catalog is stable and no more changes occur.
\end{enumerate}

This correction strategy greatly enhances our group finder's effectiveness, especially for large groups. The precision of central galaxy allocation increases from 82\% to approximately 90\%. 

\section{Results}

In this section, we perform a quantitative assessment of our group finder's performance on all the test datasets described in Section \ref{test}, which include: 
\begin{enumerate} 

\item Basic test datasets 

\item Extended test datasets: different magnitude limit  dataset, high-z dataset, and TNG dataset 

\item Readshift space dataset \end{enumerate}

To assess performance, it is crucial to align the identified groups (IGs) discovered by the group finder with the actual groups (TGs) in the simulation datasets. However, due to unavoidable inaccuracies in assigning member galaxies, the IGs and TGs may not have identical membership. According to \citet{campbell2015assessing}, these inaccuracies usually lead to two types of failure: 'fracturing' and 'fusing.' Fracturing happens when galaxies belonging to one true group are mistakenly split into multiple identified groups, whereas fusing occurs when galaxies from separate true groups are erroneously combined into one. These failure modes can appear separately or together, complicating the alignment of IGs and TGs. For clarity in presentation, the following notation will be employed:

\begin{itemize}
    \item IG: A group identified by the group finder
    \item TG: A true group within the simulation data, located in a host halo
    \item IG-T: An identified group matched to a true group
    \item TG-I: A true group matched to an identified group
\end{itemize}

Following the method proposed by \citet{wang2020identifying}, we execute a fusion of Member Matching and Central Matching to align IGs with TGs. Member Matching occurs when over 50\% of an IG's members are also within a TG, and vice versa. Central Matching is achieved if the primary galaxy of an IG coincides with that of a TG. Typically, Member Matching is regarded as more reliable, though it is more stringent for smaller assemblages. The combined approach seeks IGs that fit both member and central matching criteria for a TG, and if the outcomes of these methods diverge, we give preference to the Member Matching result. This matching procedure produces a collection of one-to-one pairings, as previously defined as IG-Ts and TG-Is.

We evaluate the model's effectiveness using the corresponding pairs of TGs and IGs by analyzing group completeness and purity, the completeness and purity of member galaxies in groups, as well as the precision of halo mass forecasts.

\subsection{Basic test datasets}

\begin{figure*}
    \centering
    \includegraphics[width=\textwidth]{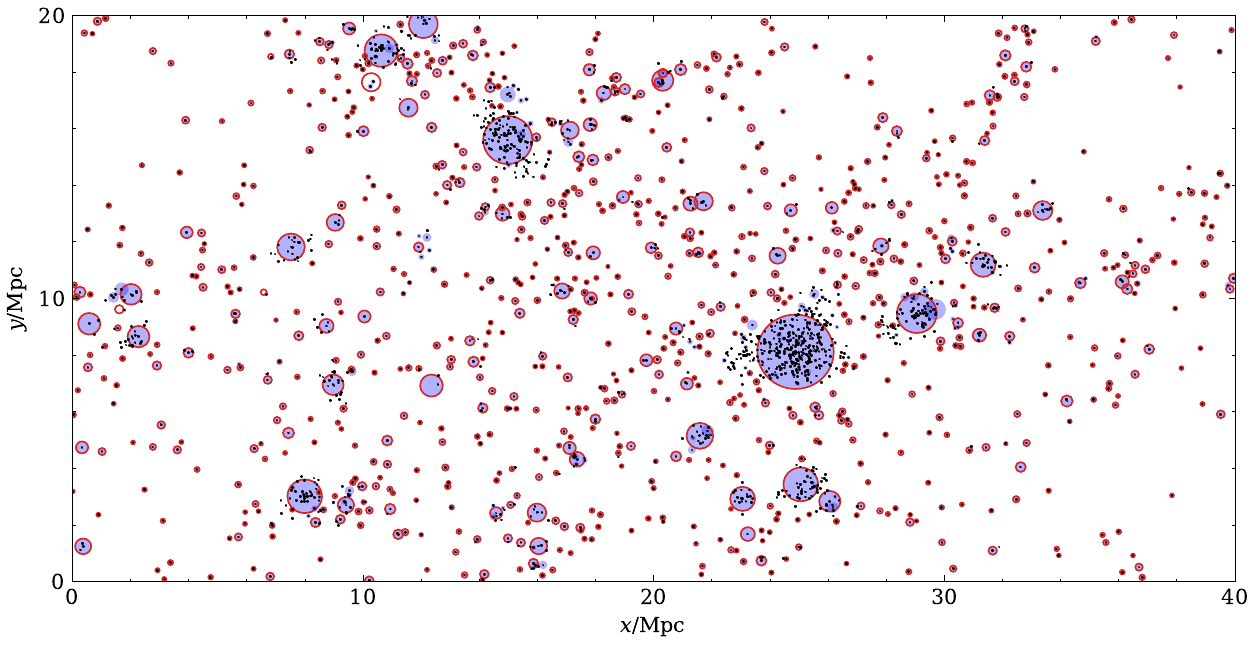}
    \caption{Comparison of true and predicted groups within a $40 \times 20 \times 20 (\hMpc)^3$ slice in the test box. There are 2087 galaxies in this region, shown as black point. These galaxies are resided in 1,020 true groups (halos) in the simulation, represented as background blue circles, with $\mathrm{R}_{vir}$ of true groups as their radii. The groups identified by our group finder are shown as red circles, with $\mathrm{R}_{vir}$ (calculated from predicted $\mathrm{M}_{vir}$) as radii. The identified groups and true groups show strong agreement in both position and size. For example, for the largest halo, we successfully assigned 316 out of 355 true member galaxies and predicted a virial mass of $10^{14.57} \hMass$, with an error of less than 0.2 dex compared to the true value of $10^{14.72}  \hMass$.}.
    \label{fig: largescale}
\end{figure*}

Initially, we assessed the effectiveness of our group finder using the basic test datasets outlined in Section \ref{test}. Prior to performing quantitative analyses, we chose a particular section within the simulated test box, which corresponds to a slice measuring $60 \times 60 \times 20 (\hMpc)^3$. This section was selected to visually compare the actual groups in the simulation with those predicted by our method. In Figure \ref{fig: largescale}, blue circles denote the true groups, while red circles represent the predicted groups in this region, with $\mathrm{R}_{vir}$ as their radii. The proximity in their locations and sizes indicates the proficiency of our group finder.

\subsubsection{Group completeness and purity}\label{def:groupCP}

We employ global completeness and purity to assess the model's performance at the group level. Group completeness, represented as $C = \frac{\mathrm{N(TG-Is)}}{\mathrm{N(TGs)}}$, is the ratio of true groups (TGs) that are correctly identified by the group finder. Group purity, indicated by $P = \frac{\mathrm{N(IG-Ts)}}{\mathrm{N(IGs)}}$, shows the percentage of identified groups (IGs) accurately matched with TGs. It's crucial to note that N(TG-Is) is equal to N(IG-Ts). 

\begin{figure}
    \centering
    \includegraphics[width=\columnwidth]{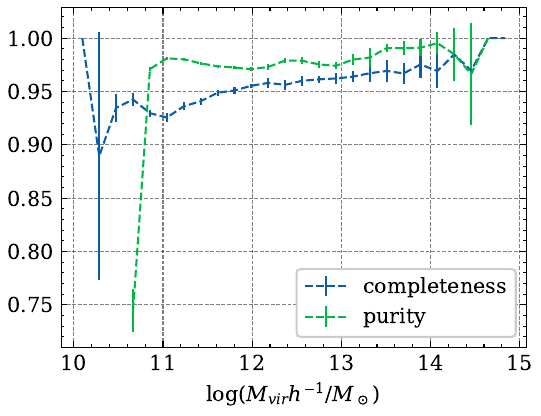}
    \caption{Group completeness and purity on the basic test datasets. See the section 4.1.1 for the definitions of group completeness and purity. The line and error bar show mean and $1\sigma$ value across the six test datasets. Completeness is shown in blue as a function of the virial mass of true groups $\mathrm{M}_{vir, \mathrm{true}}$, while purity is shown in green as a function of the predicted virial mass of identified groups $\mathrm{M}_{vir, \mathrm{predict}}$. All Groups with $\mathrm{M}_{vir} \geq 10^{11}\hMass$ achieve greater than $90\%$ completeness and purity.}
    \label{fig:groupCP}
\end{figure}

Figure \ref{fig:groupCP} illustrates the group completeness and purity for the basic test datasets. We plot how group completeness varies with the virial mass of the true groups $\mathrm{M}_{vir, \mathrm{true}}$ and how group purity changes with the group mass predicted by our group finder $\mathrm{M}_{vir, \mathrm{predict}}$. The blue line denotes the average completeness over the six basic test datasets, with error bars showing the $1\sigma$ deviation. The green line and error bars similarly depict the average purity value and standard deviation for the six datasets. Both group completeness and purity tend towards 100\% as virial mass increases and maintain a level above 90\% for all groups with $\mathrm{M}_{vir} \geq 10^{11}\hMass$. Notably, purity consistently exceeds 95\% within this mass range.

The findings reveal that our group finder successfully detects most actual halos, and a significant percentage of the discovered groups align with true groups. It is worth noting that completeness is somewhat diminished for low-mass halos. This reduction is mainly due to the increased likelihood of smaller halos being mistakenly classified as extensions of adjacent, larger halos, which aligns with the 'fusing' error category. Nonetheless, low-mass groups are uncommon and incomplete in both our test datasets and real-world surveys, so they do not substantially affect the overall efficacy of the group finder.

\subsubsection{Group member completeness and purity}

We further assess the group finder's effectiveness in assigning galaxies as group members. Although member and center matching assists in recognizing groups, the precision in determining whether the assigned galaxies truly belong to their host groups is uncertain. Consequently, we establish metrics to evaluate performance within the groups.

Consider a galaxy group identified by our group finder, which comprises N$_i$ predicted member galaxies. Assume that its corresponding true group (halo) contains N$_t$ member galaxies. If N$_s$ galaxies are shared between the true and predicted members, we define the following metrics:

\begin{enumerate}
    \item Member Completeness: $f_c = \frac{\mathrm{N}_s}{\mathrm{N}_t}$
    \item Member Purity: $f_p = \frac{\mathrm{N}_s}{\mathrm{N}_i}$
\end{enumerate}

These metrics can solely be determined for matched pairs of actual and predicted groups, since defining N$_t$ and N$_s$ is infeasible for unmatched groups. The percentage of matched identified groups corresponds to group purity, as mentioned in Section \ref{def:groupCP}.

\begin{figure*}
    \centering
    \includegraphics[width=0.9\textwidth]{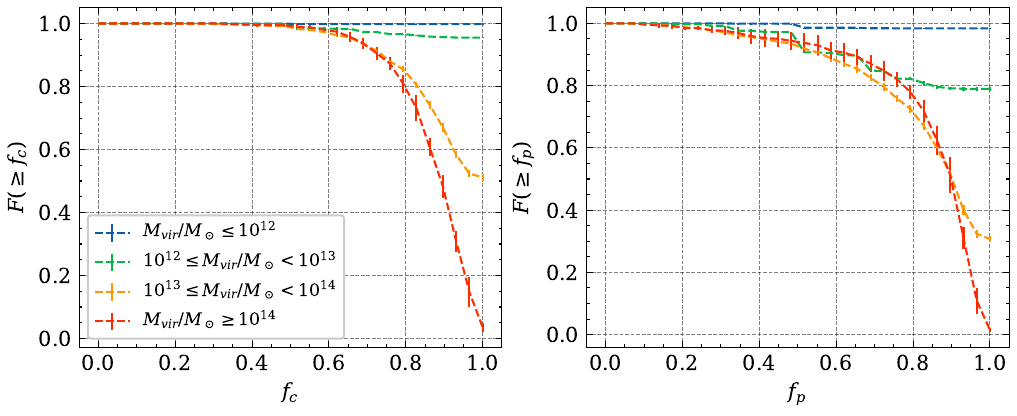}
    \caption{Member completeness and purity on the test datasets.
\textit{Left panel}: The x-axis represents the member completeness of groups ($f_c$), while the y-axis indicates the fraction of galaxies with a completeness exceeding $x=f_c$. The results are presented for 4 mass bins, each distinguished by different colors. The lines and error bars show mean and 1$\sigma$ value of the results across four mass bins.
\textit{Right panel}: Similar to the left panel, but shows the result of member purity of groups.}
    \label{fig:memberCP}
\end{figure*}

Figure \ref{fig:memberCP} presents the member completeness and purity for the test datasets. The left panel shows the cumulative distribution of member completeness ($f_c$), indicating the proportion of groups with a completeness of at least $x=f_c$. Different line styles correspond to four specific mass bins within the test sample. 

Notably, nearly all low-mass groups ($\mathrm{M}_{vir} < 10^{12}\hMass$) reach a completeness of $f_c = 1$, highlighting the model's proficiency in galaxy membership allocation. The smaller number of galaxies in low-mass groups makes full member identification easier. For groups with masses from $10^{12}\hMass$ to slightly under $10^{13}\hMass$, there is a minor decline in member completeness ($f_c$), yet approximately 95\% still achieve $f_c = 1$. Conversely, high-mass groups, due to their larger number of members, often miss some members, especially those at the edges. Approximately 80\% of high-mass groups ($\mathrm{M}_{vir} > 10^{13}\hMass$) reach a completeness of 0.8, with around 50\% to 70\% achieving $f_c \geq 0.9$. 

The right panel depicts the cumulative distribution of member purity ($f_p$), showing the fraction of groups with $f_p$ at least $x=f_p$. Like completeness, member purity for low-mass groups is almost 1. For mid-mass groups, $f_p$ is slightly lower than $f_c$. In high-mass groups, roughly 80\% achieve $f_p \geq 0.8$, with 50\% reaching $f_p \geq 0.9$. Overall, the membership evaluation suggests that our group finder efficiently manages most member assignments.

\subsubsection{Halo mass}

Determining the masses of galaxy groups is essential for compiling a catalog of such groups. In training the group mass estimator, we input the characteristics of genuine groups into the neural network, yet the model is designed to predict masses for more than just genuine groups. At each iteration step of group-finding process, a predicted mass is assigned to every potential group, significantly influencing the merging process of these potential groups. The previously demonstrated completeness and purity of groups, along with their member galaxies, indicate that the mass estimator is adept at predicting the masses of candidate groups in the iterative correction steps as well. In this context, we concentrate on the mass distribution of the final results produced by the group finder.

\begin{figure}
    \centering
    \includegraphics[width=\columnwidth]{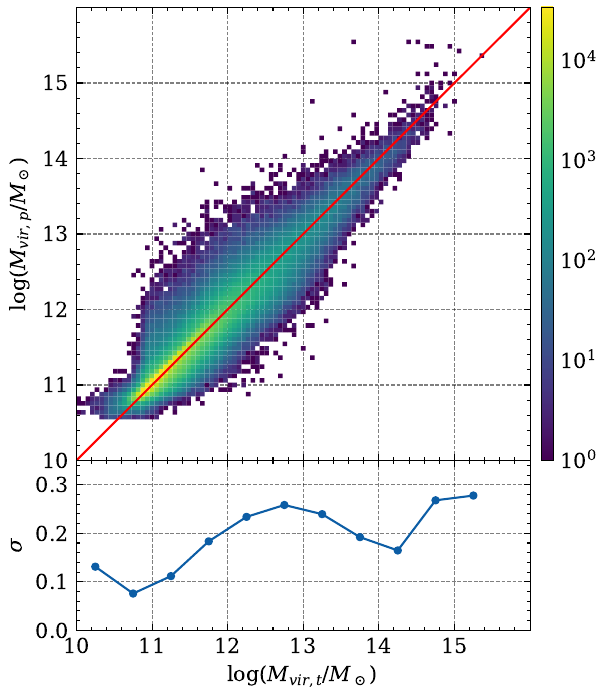}
    \caption{Comparison of true and predicted group mass for the basic test dataset. The upper panel shows a 2D histogram comparing the true and predicted masses. The x-axis represents the virial mass of true groups in simulation, while the y-axis represents the predicted mass of groups identified by group finder. Both true and predicted mass are divided into 100 bins across the mass range of $10^{10}\hMass$ to $10^{15.5}\hMass$. The red solid line represents the ideal 1:1 relation. In the lower panel, we show the standard deviation of $(\log \mathrm{M}_{vir, \mathrm{true}} - \log \mathrm{M}_{vir, \mathrm{predict}})$.}
    \label{fig:predictGroupMass}
\end{figure}

Our group finder is capable of assigning a mass to each predicted group based solely on observable properties of its member galaxies. Figure \ref{fig:predictGroupMass} shows a comparison between the actual mass of the groups and the estimated mass of the detected ones. This analysis, akin to determining the completeness and purity of the group members, is feasible only for pairs of true and identified groups that have been accurately paired, as both $\mathrm{M}_{vir, \mathrm{true}}$ and $\mathrm{M}_{vir, \mathrm{predict}}$ are necessary. The proportion of these accurately matched groups indicates the groups' completeness and purity, illustrated in Figure \ref{fig:groupCP}. The variance in the forecasted group mass is below 0.3 dex across all mass bins. The standard deviation tends to be slightly higher for mass ranges of $10^{12}\hMass \sim 10^{13}\hMass$ and $>10^{14.5}\hMass$. The first rise is attributed to the stellar mass-to-halo mass association, introducing more variability within the $10^{12}\hMass$ to $10^{13}\hMass$ span. The second peak appears at the largest masses, likely due to limited training data for such substantial halo masses.

\begin{figure}
    \centering
    \includegraphics[width=\columnwidth]{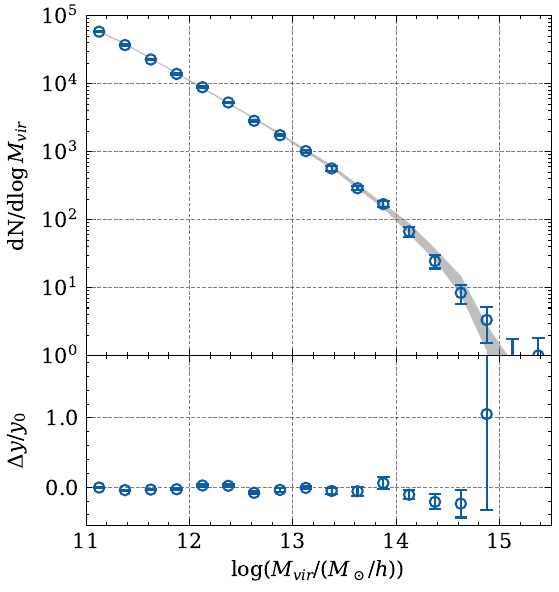}
    \caption{Comparison of predicted and true halo mass functions in the basic test datasets. The halo mass range $[10^{11}\hMass, 10^{15.5}\hMass]$ is divided into 50 bins. The grey area shows the $1\sigma$ range of the true mass distribution across all six basic test datasets. Blue points and error bars represent the predicted mean and $1\sigma$ results for the test datasets. The two functions show only slight differences for halo masses $\geq 10^{14.5} M_{\odot}$, which may be due to the limited number of large groups.}
    \label{fig:msi_hmf}
\end{figure}

Figure \ref{fig:msi_hmf} not only provides a direct comparison between the masses of true and identified groups but also displays the halo mass functions for all groups, including unmatched ones. The simulation's true halo mass function is depicted as a shaded grey region, marking the $1\sigma$ interval across the six basic test datasets. In contrast, the group finder's predicted halo mass function is illustrated with blue points and error bars, showing the mean and $1\sigma$ values derived from the six test sets. Across the entire four-order mass range, both functions are largely consistent with error less than 20\%,  except for the largest group. This discrepancy maybe arises due to the scarcity of these large groups.

\subsection{Results for extended test datasets}\label{sec:extended test}

While our model is developed and validated using simulated galaxy catalogs, we are confident that this methodology can also be applied to real observational data. For flexibility, our technique uses fundamental input features and employs a straightforward neural network architecture, which aids in reducing overfitting to the training data. To evaluate the model's performance with different real galaxy surveys, we applied the comprehensive datasets described in Section \ref{sec:extended test} to test the group's identification efficiency. Details of these datasets are provided as follows:

\begin{enumerate}
    \item Magnitude limited datasets: These samples is similar to those found in the initial test datasets, all sourced from the Millennium Simulation at a redshift of 0. They differ by employing distinct r-band apparent magnitude cutoffs: $mag_r < 16$, $mag_r < 15$, and $mag_r < 14$. This variation enables an evaluation of the model's effectiveness for shallower surveys.
    
    \item High-z datasets. These datasets comprise galaxy samples extracted from higher-redshift snapshots within the Millennium Simulation, notably at redshifts $z = 0.32$ (Snapshot 52), $z = 0.62$ (Snapshot 46), and $z = 1.08$ (Snapshot 40). The apparent magnitude constraints are consistent with those found in the fundamental test datasets. These samples assist in assessing whether our model is affected by the evolution of galaxies and halos.
    
    \item TNG300 datasets. The datasets are derived from the TNG300 simulation at redshift 0, using identical apparent magnitude limits as those in the training dataset. This allows for an evaluation of how our scheme depends on the galaxies' physical models and cosmological parameters.

\end{enumerate}

Table \ref{tab:extest} shows the fundamental characteristics of galaxies, dark matter halos, and galaxy groups as estimated by the model for these datasets. The model exhibits robust predictive accuracy across these varied samples, which are notably different from the training and basic test datasets. Group completeness stays around or surpasses 90\%, and group purity remains consistently over 97\%. Within the extended datasets, the TNG300 sample achieves the most outstanding overall results.

\begin{table*}
    \centering
    \begin{tabular}{|l|r|r|r|r|r|}
        \hline
        Catalog & Total galaxies & Total halos & Total groups & Group completeness & Group purity \\
        \hline
        MSI\_Mag16 & 169,375 & 125,850 & 120,295 & 0.93 & 0.98 \\
        MSI\_Mag15 & 118,931 & 89,325 & 86,077 & 0.94 & 0.97 \\
        MSI\_Mag14 & 77,950 & 58,675 & 56,774 & 0.94 & 0.97 \\
        MSI\_z0.32 & 593,102 & 421,840 & 396,798 & 0.92 & 0.98 \\
        MSI\_z0.62 & 651,365 & 458,889 & 406,857 & 0.87 & 0.98 \\
        MSI\_z1.08 & 670,209 & 470,646 & 423,472 & 0.88 & 0.98 \\
        TNG300 & 338,161 & 221,971 & 217,755 & 0.95 & 0.97 \\
        \hline
    \end{tabular}
    \caption{Performance of the group finder across various catalogs. Each entry includes the total counts of galaxies, halos (true groups), and predicted groups, as well as group completeness and purity for evaluating the model's effectiveness. The results illustrate the model’s robust performance, even when applied to diverse datasets beyond its original training set}
    \label{tab:extest}
\end{table*}

\begin{figure*}
    \centering
    \includegraphics[width=0.9\textwidth]{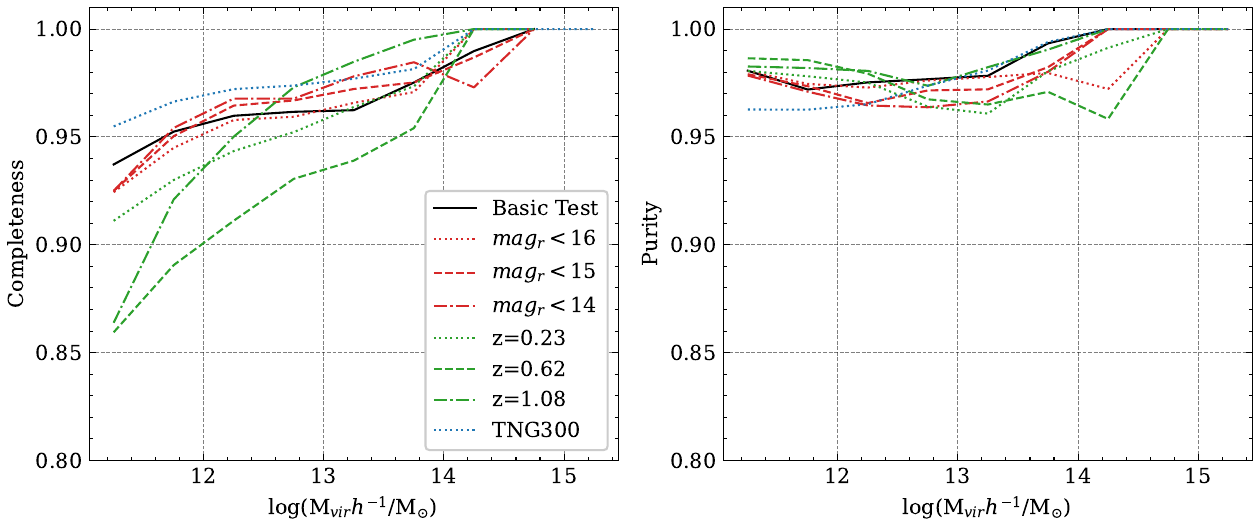}
    \caption{Completeness and purity of galaxy groups for the extended test datasets. Red lines represent the results for different magnitude limited datasets: the dotted line corresponds to the sample with $mag_r \leq 16$, the dashed line to $mag_r \leq 15$, and the dashdot line to $mag_r \leq 14$. Green lines represent high-z dataset, with the dotted line for $z=0.32$, the dashed line for $z=0.62$, and the dashdot line for $z=1.08$. The blue dotted line represents the TNG300 dataset, while the solid black line serves as a benchmark for the basic test datasets' results. Except for the low-mass intervals in $z=0.62$ and $z=1.08$, completeness for all samples exceeds 90\%, and purity exceeds 95\% for all samples.}
    \label{fig:extest_CP}
\end{figure*}

Figure \ref{fig:extest_CP} offers a detailed portrayal of the concepts of group completeness and purity. It shows the variations in completeness and purity across different mass ranges for each extended test dataset. For reference, the basic test datasets are also shown (black solid line). At redshifts $z=0.62$ and $z=1.08$, usually characterized by lower completeness, values drop below 90\% only in the lowest mass range (below $10^{12}\hMass$). Notably, some datasets demonstrate greater group completeness than the basic test data. The TNG300 dataset, in particular, maintains a completeness rate over 95\% across all mass categories. 

In terms of group purity, each sample consistently surpasses 95\% in every mass range, closely aligning with the basic test set results. This serves as evidence for the rarity of the model incorrectly dividing a single true group into multiple predicted groups, known as the fracturing failure mode. Furthermore, there is no direct linear relation between group completeness, group purity, apparent magnitude thresholds, and redshift; these measures display varied patterns across different mass ranges.

\begin{figure*}
    \centering
    \includegraphics[width=0.9\textwidth]{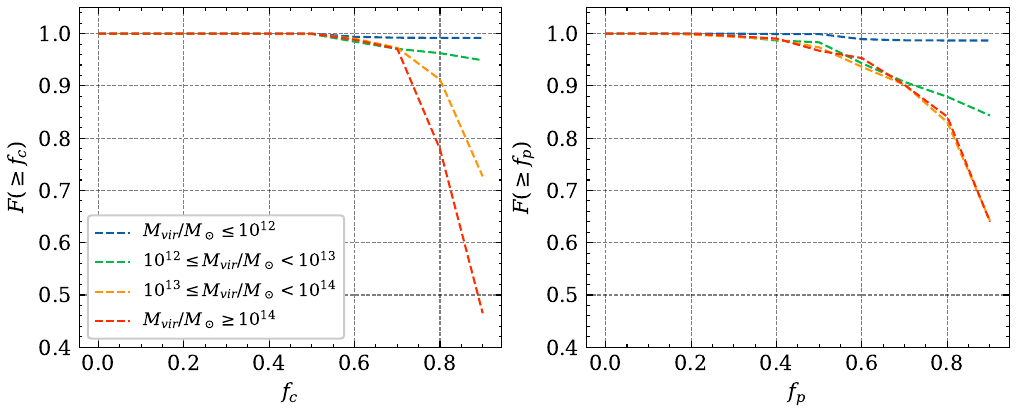}
    \caption{Completeness and purity of member galaxies in the TNG300 sample. Same as Fig \ref{fig:memberCP}, change dataset to TNG300. The results exhibit similar or even better performance compared to the MSI test set. More than 80\% of galaxy groups have completeness and purity exceeding 0.8, and over 70\% of galaxy groups have completeness and purity exceeding 0.9.}
    \label{fig:extest_member_tng}
\end{figure*}

Concerning the completeness and purity of the member galaxies ($f_c$ and $f_p$), the extended test datasets yielded quite positive outcomes. Figure \ref{fig:extest_member_tng} illustrates findings for the TNG300 dataset, which emerges as the most exemplary among the expanded test datasets. The values of $f_c$ and $f_p$ in the TNG300 sample align closely with the original test data, with over 80\% of galaxy groups achieving $f_c, f_p \geq 0.8$. Considering that our model was developed using the Millennium Simulation (a semi-analytic simulation), while TNG300 is a hydrodynamical model, the disparities in physical processes, simulation parameters, and cosmological parameters between the two are substantial. These outcomes highlight the strong generalizability of our model across different simulation datasets. This offers a promising basis for adapting our model to real observational data.

\begin{figure*}
    \centering
    \begin{subfigure}[]{\textwidth}
        \centering
        \includegraphics[width=0.9\textwidth]{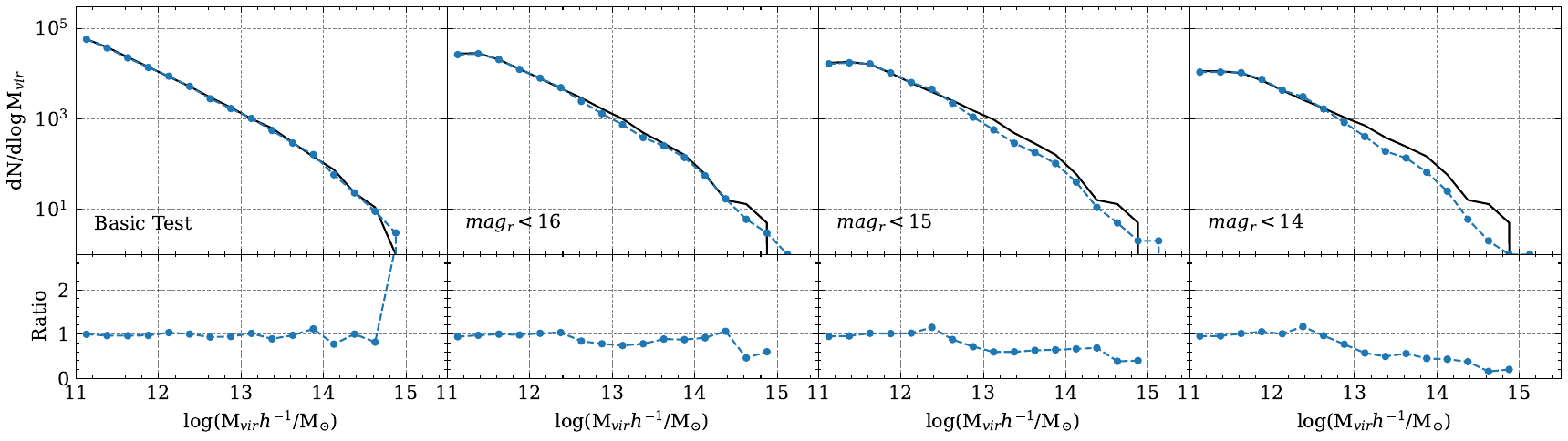}
    \end{subfigure}
    \begin{subfigure}[]{\textwidth}
        \centering
        \includegraphics[width=0.9\textwidth]{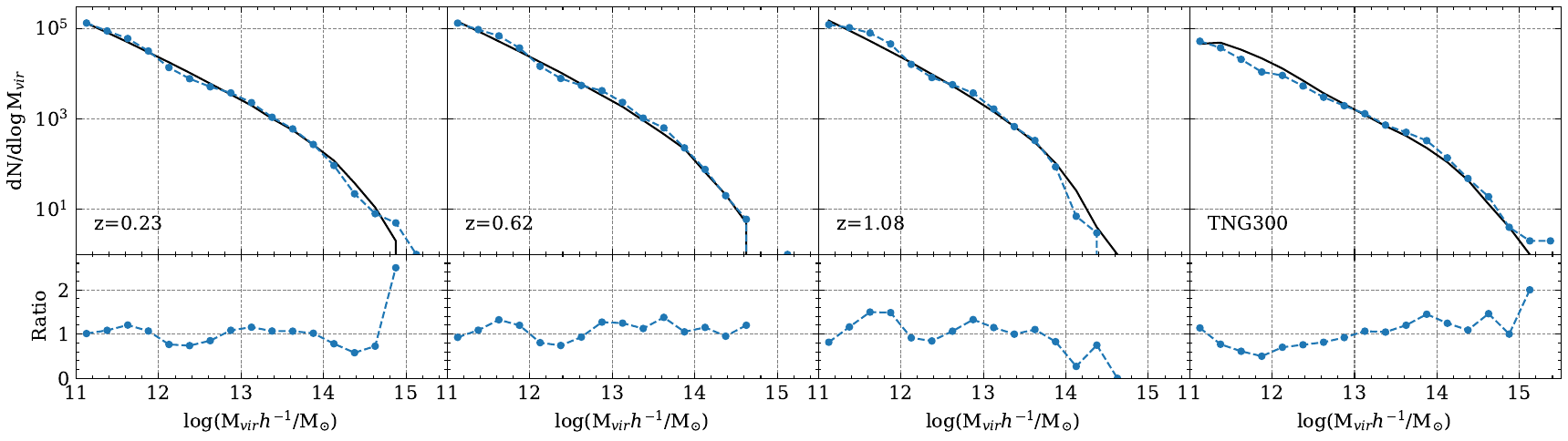}
    \end{subfigure}
    \caption{Compression of predicted halo mass function and true halo mass function on basic test set and extended test set. each column of panels show one of the  dataset. The black solid line in each upper panel indicates the halo mass function directly from simulation, and blue dashed line shows the predicted one. Each lower panel represents how the corresponding ratio of halo mass distribution changing with mass.}
    \label{fig:extest_hmf}
\end{figure*}

Evaluating predicted group masses is another crucial facet of the galaxy group catalogue. Figure \ref{fig:extest_hmf} illustrates the halo mass function (HMF) for the base dataset and seven extended datasets. The upper panel contrasts the actual HMF (black solid line) with the estimated HMF (blue dotted line), while the lower panel displays the ratio of these distributions over varying mass intervals. While the outcomes for the extended datasets aren't as precise as those for the base dataset, the predicted halo mass functions still largely align with the actual data. Minor discrepancies arise from intrinsic differences between the base and extended datasets. We will discuss these differences and their effects in detail.

\begin{figure}
    \centering
    \includegraphics[width=\columnwidth]{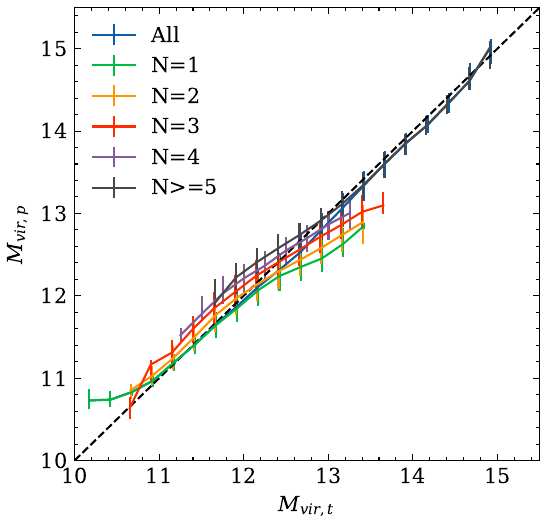}
    \caption{Dependence of group richness on Group Mass Estimator model. The figure displays the median and 1-sigma errors of predicted group mass for various true mass bins. Different colors represent galaxy groups with different richness, where blue represents all galaxy groups.}
    \label{fig:extest_massmodel_N}
\end{figure}

\begin{figure}
    \centering
    \includegraphics[width=\columnwidth]{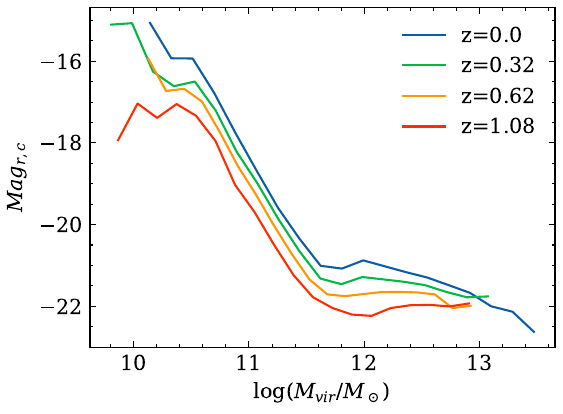}
    \caption{Comparison of the relation between halo mass and $Mag_r$ of central galaxy. It presents the curve of mean $Mag_r$ as a function of halo mass for four different snapshots. An evident linear bias is observable.}
    \label{fig:extest_highz}
\end{figure}

\begin{figure}
    \centering
    \includegraphics[width=\columnwidth]{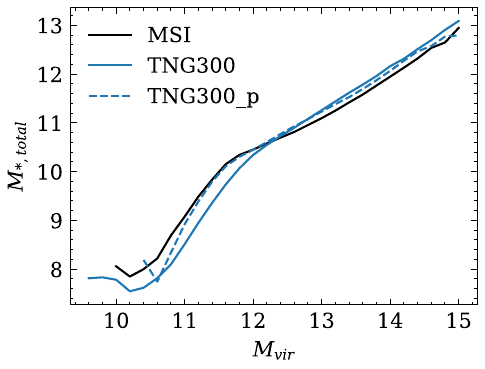}
    \caption{Comparison of Stellar Mass - Halo Mass relation on MSI and TNG300. The solid black line represents the median curve of total stellar mass corresponding to the dark halo mass in MSI, while the solid blue line represents the relation within the TNG300 sample. There is a noticeable discrepancy between the two. The blue dashed line also from TNG300 but using model-predicted mass, and it becomes closer to MSI's results.}
    \label{fig:extest_tng}
\end{figure}

In the magnitude limited datasets, the mass distribution function aligns well with actual values for lower masses (below $10^{12.5}$). However, deviations become noticeable in the higher mass range, escalating as the apparent magnitude threshold is decreased. Adjusting this limit affects several input parameters for the mass estimation models, including $\mathrm{M}_{*, \mathrm{true}}$, $Mag_{\mathrm{max}}$, and properties of satellite galaxies. Analysis of these distributions and their significance revealed that group richness $N$ is the most influential factor. In groups with more than five satellite galaxies, raising the apparent magnitude limit primarily affects $N$, while most other parameters remain stable. Additional data analysis indicates performance variations based on the galaxy clusters' member richness. Figure \ref{fig:extest_massmodel_N} presents the mass prediction results for basic test datasets over various halo mass ranges, highlighting the median predicted values with corresponding 1-sigma errors. We classified the samples into five groups based on the number of member galaxies: $N=1$, $N=2$, $N=3$, $N=4$, and $N \geq 5$. The figure demonstrates that, for groups sharing the same halo mass, those with more member galaxies tend to have higher predicted masses. This causes slight underestimation of masses for groups with fewer members, especially for larger groups. The increasing presence of such groups in samples with higher apparent magnitudes contributes to a decline in the mass distribution function's high-mass end. Notably, the strictness of the $mag_r < 14$ limit renders it unlikely to be used in modern redshift surveys.

In high-redshift datasets, the calculated halo mass function aligns reasonably well with the actual data, although discrepancies grow as redshift increases. We categorized the samples into two groups: those with $N>1$ and those with $N=1$. The primary discrepancies in mass predictions are found in the galaxy groups with $N=1$. Out of all the input parameters for mass prediction, only $M_c$ and $Mag_{r, c}$ (equivalent to $=Mag_{max}$) show effectiveness for these particular groups. Upon examining their correlation with halo mass, we discovered that this relation fluctuates significantly across different redshifts. Figure \ref{fig:extest_highz} displays the link between the median $Mag_{r, c}$ and halo mass, where curves for various redshifts show a unique, nearly linear bias. For the z=1.08 samples compared to the test datasets, there's an approximate deviation of 1 magnitude. A straightforward correction was tested by adding 1 to $Mag_{r, c}$ for each sample within the z=1.08 dataset, leading to a median mass prediction closely matching the actual mass. This indicates that applying a linear adjustment to the magnitudes of galaxies at higher redshifts can maintain our model's relevance for galaxies across varying redshifts.

The TNG300 dataset is notably distinct from the training data and other samples as it is derived from hydro-dynamic simulations and employs a different cosmological model. Despite this difference, the predicted halo mass function maintains an accuracy comparable to high-redshift datasets. However, there is a minor bias: smaller groups tend to have their masses underestimated, whereas larger groups see an overestimation. This variance is linked to inherent disparities in the Stellar Mass - Halo Mass (SMHM) relation between the TNG300 and Millennium simulations. Figure \ref{fig:extest_tng} demonstrates the correlation between halo mass and average stellar mass within both simulations. A notable difference is evident: for lower mass ranges, TNG300 forecasts a lower total stellar mass for a given halo mass, whereas for higher mass ranges, it anticipates a greater stellar mass. The figure's blue dashed line portrays the connection between predicted halo mass and average stellar mass, aligning closely with the Millennium Simulation's results. It matches the Millennium Simulation's outcomes at the low mass end because these groups are predominantly isolated centers or have only few satellites, implying that the prediction aligns with the Millennium SMHM relation. However, as the number of satellites increases, the predicted halo mass aligns more accurately with the true value in TNG300, rather than adhering strictly to the relation in the training data. This highlights the network's capability in understanding the link between halo mass and the properties of central and satellite galaxies fed into it. Expanding our network's application to various galaxy samples, including those from actual galaxy surveys, is anticipated to produce accurate predictions for halo mass.

These extended test set results demonstrate that our model is capable of generating dependable outcomes without needing to retrain on various mock catalogs. Group completeness is generally around 90\%, and purity consistently surpasses 95\% for these datasets. The completeness and purity of member identification are also high, confirming the model's efficacy in detecting group members. Additionally, the model can reliably reconstruct the halo mass function. However, small biases in estimating the halo mass of galaxy clusters may occur due to inherent differences between the training/basic data and the  extended datasets.

\subsection{Redshift distorted samples}

While our model is constructed for an optimal setting, we propose that it can be successfully adapted to more practical situations. This adaptability is due to its limited dependence on exact redshift measurements used for network training, as outlined in the Method section. In identifying nearby targets, we consistently incorporate a $\pm 5\hMpc$ margin of uncertainty in line-of-sight distance measurements. This approach is utilized when locating the nearest 10 neighbors of target galaxies and when looking for all galaxies within $R_{200}$ of a potential group. The first try we should do is to extend the test into the redshift distorted sample. 

In this test, we preserve the essential structure of our model but re-train it using a simulated redshift distorted sample. This catalog is modeled on galaxy data from the $z = 0$ snapshot of the Millennium Simulation, incorporating redshift distortions based on the galaxies' velocities along a specified axis. We apply the same apparent magnitude threshold to these redshift distorted samples as utilized for the test data. 

\begin{figure*}
    \centering
    \begin{subfigure}[b]{0.45\textwidth}
        \centering
        \includegraphics[width=0.8\textwidth]{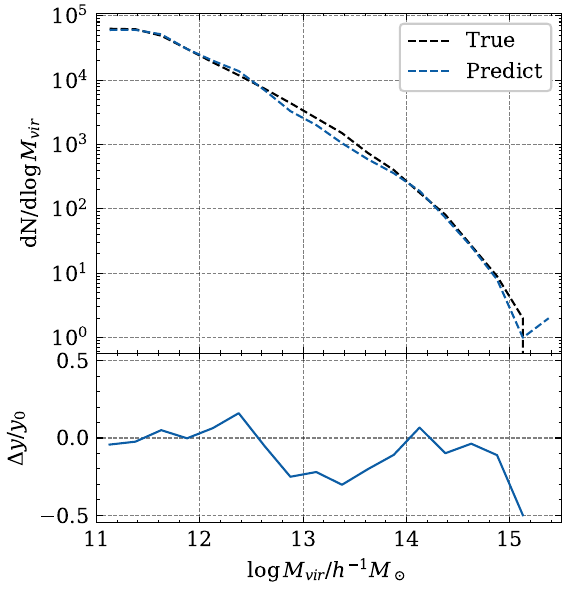}
        \caption{}
        \label{fig:mock_hmf}
    \end{subfigure}
    % \hfill
    \begin{subfigure}[b]{0.45\textwidth}
        \centering
        \includegraphics[width=\textwidth]{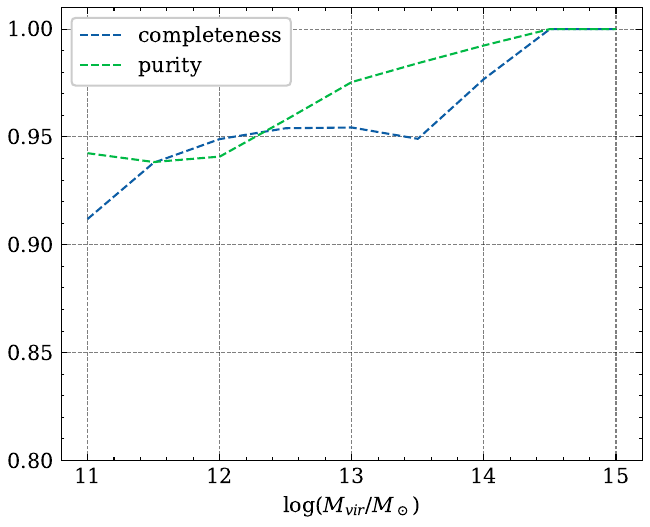}
        \caption{}
        \label{fig:mock_groupcp}
    \end{subfigure}

    \begin{subfigure}[b]{0.9\textwidth}
        \centering
        \includegraphics[width=\textwidth]{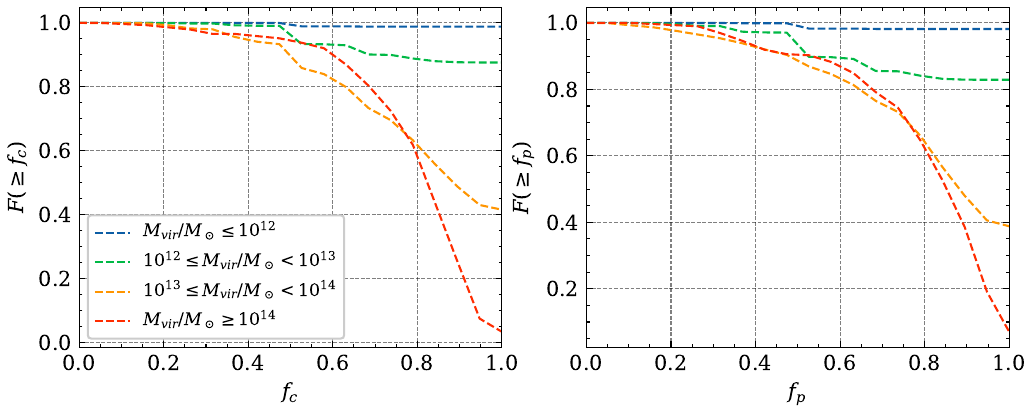}
        \caption{}
        \label{fig:mock_member}
    \end{subfigure}
    
    \caption{Results on redshift distorted samples. (a) Comparison of halo mass function. Black dashed line is the true value of simulation, and blue dashed line shows the prediction of our group finder. The lower panel present $(\log \mathrm{M}_{p, vir} - \log \mathrm{M}_{t, vir}) / \log \mathrm{M}_{t, vir}$ value on each mass bins. (b) The group completeness and purity in redshift distorted samples. The definition of them can be found in section \ref{def:groupCP}. We only show results for groups with $\mathrm{M}_{vir} \geq 10^{11}\hMass$ because smaller groups are highly incomplete due to selection effect. (c) The completeness and purity function of member galaxies in redshift distorted samples.} 
    \label{fig:mock_test}
\end{figure*}

Figure \ref{fig:mock_test} displays the result for redshift distorted sample. The mass estimations in Figure \ref{fig:mock_hmf} align well with true mass values for both smaller and larger mass groups, although there is a tendency to slightly underestimate in the $10^{13}\hMass \sim 10^{14}\hMass$ range. This underestimation stems mainly from the incomplete membership predictions in this mass range. Both group completeness and purity exceed 90\% across all mass classes (Figure \ref{fig:groupCP}) and surpass 95\% for groups with $\mathrm{M}_{vir} > 10^{12}\hMass$, showcasing our model's strong ability to detect groups in redshift space. In Figure \ref{fig:mock_member}, the completeness curve declines more sharply compared to the original test data, with around 60\% of groups having $M_{vir} \geq 10^{13} M_{\odot}$ achieving $f_c \geq 0.8$. The overall performance on redshift distorted samples illustrates the significant capability of our model for application in real redshift surveys. Minor predictive errors suggest the need for additional adjustments and optimizations for use in redshift space contexts. More precise results for real surveys will be presented in our subsequent papers, including group catalogs from actual redshift surveys.

\section{Conclusion}

In this research, we devised a machine learning-based strategy to identify galaxy groups by leveraging various observable characteristics of galaxies. This methodology exhibited strong performance when tested on simulated mock catalogs. Our galaxy group identification system is composed of three primary elements:

\begin{enumerate}
    \item Central Galaxy Identifier. This component is essential for distinguishing central and satellite galaxy pairs within a group by employing a machine-learning classification algorithm. It assesses a target galaxy and its ten closest neighbors to accurately identify the central galaxy with about 90\% accuracy.
    
    \item Group Mass Estimator. We developed a regression model to estimate the halo mass of galaxy groups found by our system. This model uses the attributes of a group's central galaxy and its five most massive satellites as inputs. Evaluation against a true group catalog showed a prediction error rate of approximately 0.2 dex.
    
    \item Group finder. Although the Central Galaxy Identifier achieves significant accuracy in generating a galaxy group catalog, sporadic segmentation errors, particularly in more massive groups, were observed. To mitigate these, we implemented an iterative procedure to consolidate fragmented sections of actual groups.
\end{enumerate}

The performance of the group identification tool was thoroughly assessed using six basic test datasets derived from the z=0 snapshot of the Millennium Simulation (MSI). The findings reveal that the completeness and purity of the groups surpass 90\% for all group mass ranges, including the lower mass limit near $10^{11}\hMass$. Concerning the precision of member allocation, over 80\% of the groups had member completeness above 80\%, while more than 90\% showed member purity exceeding 60\%. Additionally, the estimated halo mass distribution was in extraordinary agreement with the true values, with a ratio that approximates 1 across most mass ranges.

We conducted an additional evaluation of our group finder using three separate datasets to determine the model's extensibility and flexibility. These comprehensive test datasets comprised samples with diverse apparent magnitude thresholds, high-redshift samples, and hydrodynamic simulation TNG300 samples. 

The model reliably assigned group memberships across all datasets, achieving group completeness over 90\% and purity exceeding 95\%. Interestingly, certain datasets performed better than the standard test datasets. Nonetheless, predictions of halo mass showed some errors, especially in samples with varying selection criteria. We believe these inconsistencies may arise due to the following reasons: the prediction of our current scheme has a weak dependence on the richness of groups, this will lower the predicted virial mass for smaller groups. Also our model has a weak dependence on the Stellar Mass - Halo Mass (SMHM) relation from the training data for small groups, and then the prediction of halo mass will keep inline with the SMHM relation from the training data for smaller groups with few satellites. Furthermore, the difference between the r-band luminosity-halo mass from different cosmic epochs in real data bring extend errors on the prediction of the halo mass for distant groups.

However, for massive groups with more satellites in TNG300 sample, the predicted halo mass aligns more accurately with the true value in TNG300, rather than adhering strictly to the relation in the training data. This highlights the network's capability in understanding the link between halo mass and the properties of central and satellite galaxies fed into it. Expanding our network's application to various galaxy samples, including those from actual galaxy surveys, is anticipated to produce accurate predictions for halo mass.

Although our group finder is developed in real space, we address the impact of redshift-space distortions by setting a $\pm 5\hMpc$ uncertainty when dealing with line-of-sight distance. This approach guarantees the ability of our model in detecting galaxy groups within redshift space. Following retraining using a redshift distorted mock catalog based on the Millennium Simulation, the model consistently achieves high levels of group completeness and purity, with rates surpassing 90\% for all mass categories and exceeding 95\% for more massive groups ($M_{vir} > 10^{12}\hMass$). Minor discrepancies have been observed, specifically the underestimation of masses ranging from $10^{13}\hMass M_{\odot} \sim 10^{14}\hMass$, mainly due to incomplete member predictions in this mass range. These findings highlight the model's potential for use in real redshift surveys, although additional adjustments might be required to fully adapt it to observational data. The effectiveness demonstrated with mock catalogs underscores the prospects of our group finder for actual redshift surveys. In our subsequent studies, we intend to thoroughly examine the diverse observational effects in the actual survey, and apply our techniques to the real galaxy survey.

\begin{acknowledgements}
This work was supported by the National Key R\&D Program of China (2022YFA1602901), the NSFC grant (Nos 11988101, 11873051, 12125302, and 11903043), CAS Project for Young Scientists in Basic Research Grant (No. YSBR-062), and the K.C. Wong Education Foundation. YC acknowledges the support of the UK Royal Society through a University Research Fellowship. For the purpose of open access, the author has applied a Creative Commons Attribution (CC BY) licence to any Author Accepted Manuscript version arising from this submission.
\end{acknowledgements}

\appendix                  %%appendicial material is supported

% \section{This shows the use of appendix}
% A postscript file is actually an ASCII text file (you may even edit it).
% However, you need to transfer a PDF file or any compressed or packaged
% file in binary mode when using FTP.

% \section{What is SCI?}
% SCI is the abbreviation of Science Citation Index system powered by
% the Institute for Scientific Information (ISI). For details please
% visit {\it http://apps.isiknowledge.com}.

\bibliographystyle{raa}
\bibliography{main}

\label{lastpage}

\end{document}